\documentclass[aps,prl,twocolumn,superscriptaddress,showpacs,floatfix]{revtex4-1}
\usepackage{makecell}
\usepackage{graphicx}
\usepackage{dcolumn}
\usepackage{bm}
\usepackage{times}
\usepackage{color}

\begin{document}
\bibliographystyle{apsrev4-1}

\title{Quantum oscillations and electronic structure in the large-Chern number topological chiral semimetal PtGa}
\author{Sheng Xu}\thanks{These authors contributed equally to this paper}
\affiliation{Department of Physics, Renmin University of China, Beijing 100872, P. R. China}
\affiliation{Beijing Key Laboratory of Opto-electronic Functional Materials $\&$ Micro-nano Devices, Renmin University of China, Beijing 100872, P. R. China}
\author{Liqin Zhou}\thanks{These authors contributed equally to this paper}
\affiliation{Beijing National Laboratory for Condensed Matter Physics and Institute of Physics, Chinese Academy of Sciences, Beijing 100190, P. R. China}
\affiliation{CAS Center for Excellence in Topological Quantum Computation, University of Chinese Academy of Sciences, Beijing 100049, P. R. China}
\author{Xiao-Yan Wang}\thanks{These authors contributed equally to this paper}
\affiliation{Department of Physics, Renmin University of China, Beijing 100872, P. R. China}
\affiliation{Beijing Key Laboratory of Opto-electronic Functional Materials $\&$ Micro-nano Devices, Renmin University of China, Beijing 100872, P. R. China}
\author{Huan Wang}
\affiliation{Department of Physics, Renmin University of China, Beijing 100872, P. R. China}
\affiliation{Beijing Key Laboratory of Opto-electronic Functional Materials $\&$ Micro-nano Devices, Renmin University of China, Beijing 100872, P. R. China}
\author{Jun-Fa lin}
\affiliation{Department of Physics, Renmin University of China, Beijing 100872, P. R. China}
\affiliation{Beijing Key Laboratory of Opto-electronic Functional Materials $\&$ Micro-nano Devices, Renmin University of China, Beijing 100872, P. R. China}
\author{Xiang-Yu Zeng}
\affiliation{Department of Physics, Renmin University of China, Beijing 100872, P. R. China}
\affiliation{Beijing Key Laboratory of Opto-electronic Functional Materials $\&$ Micro-nano Devices, Renmin University of China, Beijing 100872, P. R. China}
\author{Peng Cheng}
\affiliation{Department of Physics, Renmin University of China, Beijing 100872, P. R. China}
\affiliation{Beijing Key Laboratory of Opto-electronic Functional Materials $\&$ Micro-nano Devices, Renmin University of China, Beijing 100872, P. R. China}
\author{Hongming Weng}
\affiliation{Beijing National Laboratory for Condensed Matter Physics and Institute of Physics, Chinese Academy of Sciences, Beijing 100190, P. R. China}
\affiliation{CAS Center for Excellence in Topological Quantum Computation, University of Chinese Academy of Sciences, Beijing 100049, P. R. China}
\affiliation{Songshan Lake Materials Laboratory, Dongguan, Guangdong 523808, P. R. China}
\author{Tian-Long Xia}\email{tlxia@ruc.edu.cn}
\affiliation{Department of Physics, Renmin University of China, Beijing 100872, P. R. China}
\affiliation{Beijing Key Laboratory of Opto-electronic Functional Materials $\&$ Micro-nano Devices, Renmin University of China, Beijing 100872, P. R. China}

\date{\today}
\begin{abstract}
We report the magnetoresistance(MR), de Haas-van Alphen (dHvA) oscillations and the electronic structures of single crystal PtGa. The large unsaturated MR is observed with the magnetic field B//[111]. Evident dHvA oscillations with B//[001] configuration have been observed, from which twelve fundamental frequencies are extracted and the spin-orbit coupling effect (SOC) induced band splitting is revealed. The light cyclotron effective masses are extracted from the fitting by the thermal damping term of the Lifshitz-Kosevich (LK) formula. Combining with the calculated frequencies from the first-principles calculations, the dHvA frequencies F$_1$/F$_3$ and F$_{11}$/F$_{12}$ are confirmed to originate from the electron pockets at $\Gamma$ and R, respectively. The first-principles calculations also reveal the existence of spin-3/2 RSW fermion and time-reversal (TR) doubling of the spin-1 excitation at $\Gamma$ and R with large Chern number $\pm4$ when SOC is included.

\end{abstract}

\maketitle
\setlength{\parindent}{1em}
\section{Introduction}

Weyl semimetals have attracted tremendous attention in the area of condensed matter physics. Plenty of novel and significant transport properties, such as the high mobility, negative longitudinal MR and three dimensional quantum Hall effect, are caused by the chiral Weyl fermions and Weyl orbits\cite{wan2011topological,huang2015observation,zhang2016signatures,liu2016evolution,xu2016observation,xu2015experimental,xu2015discovery2,huang2015weyl,xu2015discovery,3D-Hall,ISI:000455781600036}. The materials of TaAs family \cite{weng2015weyl,xu2015discovery,huang2015weyl,lv2015experimental,ISI:000360709200013,xu2015discovery2,xu2015experimental,xu2016observation,liu2016evolution,huang2015observation,zhang2016signatures,arnold2016negative} are well-known as Weyl semimetals with spin-1/2 twofold Weyl fermion. In addition, some new types of multifold Weyl fermions, named as spin-1 chiral fermion, double Weyl fermion and spin-3/2 Rartia-Schwinger-Weyl (RSW) fermion\cite{rarita1941theory,manes2012existence,xu2016type,liang2016semimetal,ezawa2016pseudospin,tang2017multiple,chang2017unconventional,ZhangTT2018,pshenay2018band}, are put forward and observed in CoSi, RhSi, RhSn, PtAl and PdGa\cite{tang2017multiple,pshenay2018band,chang2017unconventional,sanchez2019discovery,rao2019new,takane2019observation,schroter2019chiral,xu2019crystal,wu2019single,Li-RhSn,Xu-RhSn,schroter2019observation}. These materials with the space group P2$_1$3 (No. 198) hold the spin-1 excitation at $\Gamma$, double Weyl fermion at R in the first Brillouin zone with the Chern number $\pm$2, and two Fermi arcs in their surface states (SSs). The band splitting is induced when the SOC is considered. Thus, the spin-1 excitation and double Weyl fermion evolve into spin-3/2 RSW fermion and time-reversal (TR) doubling of the spin-1 excitation with the Chern number $\pm$4, respectively. Correspondingly, Fermi arcs should split and result in four observable Fermi arcs on the surface. However, the SOC is relatively weak and the band splitting is not observed in the recent angle-resolved photomission spectroscopy (ARPES) or transport studies on CoSi, RhSi and PtAl\cite{chang2017unconventional,sanchez2019discovery,rao2019new,takane2019observation,schroter2019chiral,xu2019crystal,wu2019single}. Band splitting in RhSn with stronger SOC has been observed and confirmed by our previous transport study\cite{Xu-RhSn}. However, the splitting in the bulk and surface states was not resolved in our previous (ARPES) data\cite{Li-RhSn}.

\begin{figure}[b]
	\centering
	\includegraphics[width=0.48\textwidth]{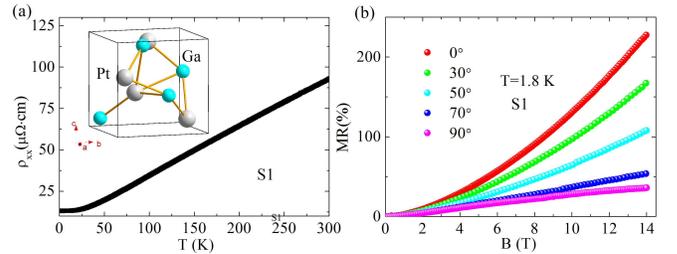}\\
\caption{(a) Temperature dependence of the resistivity $\rho_{xx}$. Inset shows the crystal structure of PtGa. (b) Magnetic field-dependent MR at 1.8 K with magnetic field titled from B$\bot$I ($\theta=0^o$) to B//I ($\theta=90^o$).}
\end{figure}

Motivated by above results and discussions, we further grew PtGa single crystals with stronger SOC and studied its transport properties and electronic structures. The crystal structure of PtGa also belongs to P2$_1$3 (No. 198). The topological properties and band structure characteristics of PtGa are believed to be similar with those in RhSn. According to our first-principles calculations, there exist spin-1 excitation at $\Gamma$ and double Weyl fermion at R in the first Brillouin zone with the Chern number $\pm$2 without considering SOC or spin-3/2 RSW fermion at  $\Gamma$ and time-reversal (TR) doubling of the spin-1 excitation at R with the Chern number $\pm$4 when SOC is considered. PtGa displays a metallic behavior in the temperature-dependent resistivity measurement and large unsaturated longitudinal MR at 1.8 K and 14 T. Evident dHvA oscillations are observed and twelve fundamental frequencies are extracted from the fast Fourier-transform (FFT) analysis. Combining with the frequencies calculated from the first-principles calculations with SOC considered, the two frequencies at 11.2 T \& 66.2 T are attributed to the electron-like pockets at  $\Gamma$  and the two frequencies at 1723.9 T \& 1937.3 T are from the electron-like pockets at R. However, according to the first-principles calculations without SOC included, there should exist only one frequency at $\Gamma$ and R, respectively. Thus, the SOC-induced band splitting is revealed by the dHvA oscillations, indicating the existence of spin-3/2 RSW fermion  at $\Gamma$ and TR doubling of the spin-1 excitation at R, respectively. The light cyclotron effective masses are extracted from the fitting of thermal damping term by the LK formula, indicating the possible existence of massless quasiparticles. PtGa provides a better platform to study the SOC-induced band splitting and split Fermi arcs in the materials of this family .

\section{Methods}
The single crystals of PtGa were grown by the Bi flux method. The platinum powder, gallium ingot and bismuth granules were put into
the crucible and sealed into a quartz tube with the ratio of
Pt:Ga:Bi=1:1:50. The quartz tube was heated to 1150 $^{\circ}$C at 60 $^{\circ}$C/h and held for 20 h, then
cooled to 550$^{\circ}$C at 1 $^{\circ}$C/h. The rod-like PtGa single crystals were obtained by centrifugation. The atomic
composition of PtGa single crystal was checked to be
Pt$:$Ga=1$:$1 by energy dispersive x-ray spectroscopy (EDS, Oxford
X-Max 50). The measurements of resistivity and magnetic properties were performed
on a Quantum Design physical property measurement system (QD
PPMS-14T).  The standard four-probe method was applied on the resistivity and magnetoresistance measurements on a long flake crystal (S1) which was cut and polished from the crystals as grown. The electrode was made by platinum wire with silver epoxy. Magnetization measurements were conducted on another smaller sample (S2) with higher quality.

The first-principles calculations of PtGa were performed by the Vienna Ab initio Simulation Package (VASP)\cite{kresse1996efficient}, which was based on the density functional theory (DFT). The generalized gradient approximation (GGA) in the Perdew–Burke–Ernzerhof (PBE) type was selected to describe the exchange-correlation functional\cite{perdew1996generalized}. The cutoff energy was set to 450 eV, and the Brillouin zone (BZ) integral for self-consistent calculation was sampled by 10 $\times$ 10 $\times$ 10 $k$ mesh.
We constructed the tight-binding model of PtGa by using the Wannier90 package with Ga 4p orbital and Pt 5d orbital, based on the maximally localized Wannier functions method (MLWF)\cite{mostofi2014updated}, and we further calculated the bulk Fermi surfaces by using the WannierTools software package\cite{wu2018wanniertools}.

\section{Results and Discussions}

The crystal structure of PtGa is illustrated in the inset of Fig. 1(a), which crystallizes in a simple cubic structure with the space group $P2_13$ (No. 198). Figure 1(a) displays the temperature-dependent resistivity which exhibits the metallic behaviour. The field-dependent longitudinal MR=($\rho_{xx}(B)-\rho_{xx}(0))/\rho_{xx}(0)$ with the magnetic field tilted from B$\bot$I  to B//I  at 1.8 K is exhibited in Fig. 1(d). The large unsaturated longitudinal MR reaches ~230\% at 1.8 K and 14 T with B//[111] configuration, which decreases gradually with $\theta$ increasing from $\theta=0^o$ to $\theta=90^o$ and tends to saturate at $\theta=90^o$ under high field. Because of the existence of trivial pockets, the negative longitudinal MR with B//I configuration in PtGa has not been observed.

\begin{figure}[t]
\centering
  \includegraphics[width=0.48\textwidth]{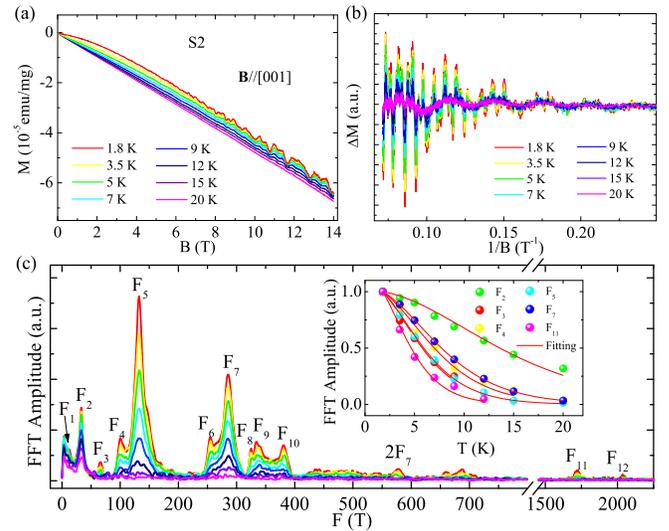}\\
\caption{(a) The dHvA oscillations at various temperatures with
\emph{B}//[001] configuration. (b) The amplitudes of dHvA oscillations as a function of 1/B. (c) The FFT spectra
of the oscillations. Inset shows the temperature dependence of relative normalized FFT amplitude of the frequencies.}
\end{figure}

\begin{figure}[h]
	\centering
	\includegraphics[width=0.48\textwidth]{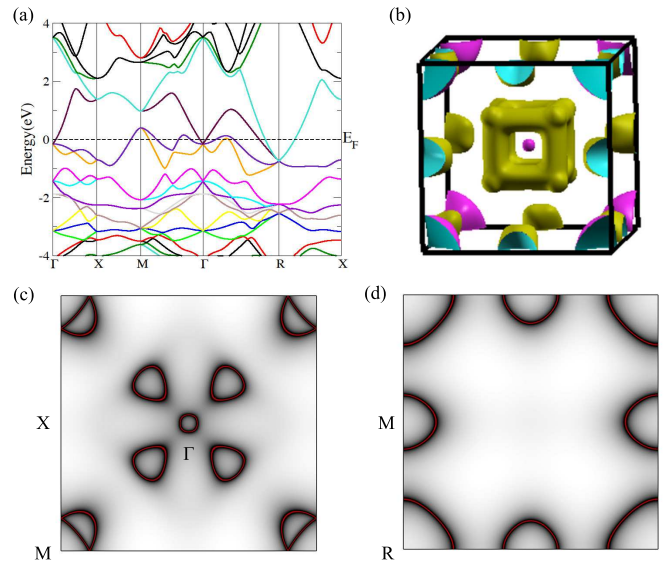}\\
	\caption{(a), (b) Calculated bulk band structure of PtGa along high-symmetry lines and FSs in the bulk Brillouin zone without SOC. (c), (d) Calculated FSs in $k_z$=0 plane and $k_z$=$\pi$ plane without SOC.}
\end{figure}

\begin{figure}[b]
	\centering
	\includegraphics[width=0.48\textwidth]{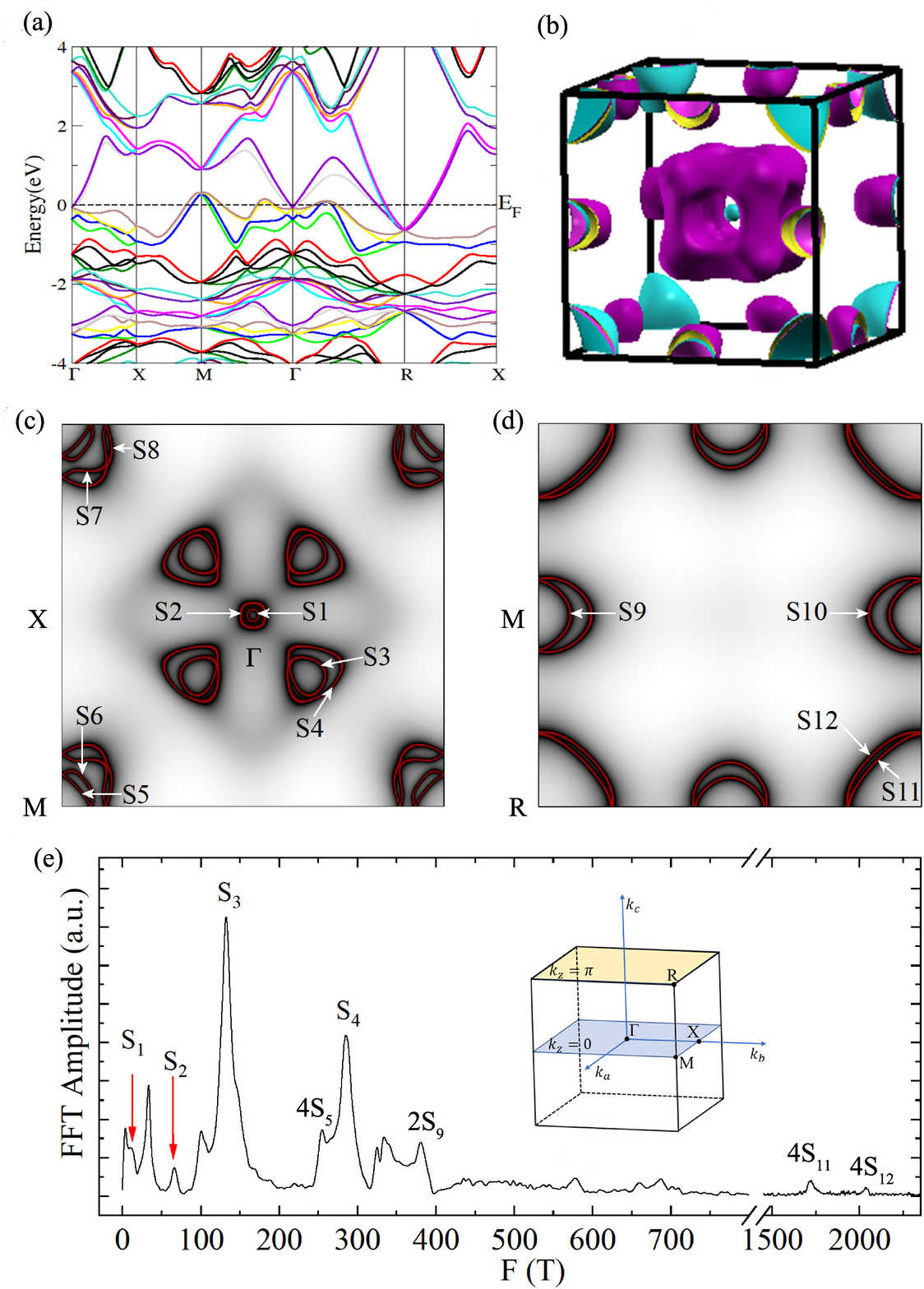}\\
	\caption{(a), (b) Calculated bulk band structure of PtGa along high-symmetry lines and FSs in the bulk Brillouin zone with SOC. (c), (d) Calculated FSs in $k_z$=0 plane and $k_z$=$\pi$ plane with SOC. (e) Matching between the FSs' extreme cross sections and the observed frequencies. Inset shows the Brillouin zone of PtGa with $k_z$=0 and $k_z$=$\pi$ planes highlighted.}
\end{figure}

\begin{table*}
  \centering
\caption{The parameters extracted from dHvA oscillations in PtGa with B//[001] configuration. F is the frequency of dHvA oscillations; $m^*$/$m_e$ is the ratio
of the effective mass to the electron mass; $\phi$$_B$ is the Berry phase. }
  \label{oscillations}
  \setlength{\tabcolsep}{2mm}{
  \begin{tabular}{ccccccccccccccc}
  \hline
  \hline
   &               &$F_1$     & $F_2$    & $F_3$ & $F_4$ & $F_5$ & $F_6$ & $F_7$ & $F_8$ & $F_9$    & $F_{10}$ &$F_{11}$ &$F_{12}$ &\\
   \hline
  &$F(T)$          &11.2      &33.2      &66.2   &100.4  &132.3   &254.7  &284.9  &324.6  &333.7    &380.4&1723.9&1937.3&\\
  &$m^*$/$m_e$     &/         &0.05      &0.12   &0.10   &0.11   &/       &0.08   &/      &/        &/    &0.15  &/&\\
  &$A_{F}$ $10^{-3}{\AA}^{-2}$
                   &1.07      &3.17      &6.32   &9.58   &12.63  &24.31  &27.20  &30.99  &31.85    &36.31&164.56&184.93&\\
  \hline
  \hline
  \end{tabular}}
  \end{table*}

The quantized Landau levels will cross the Fermi energy (E$_F$) with the increasing field, which leads to the oscillations of density of state (DOS) at E$_F$ and finally induces the quantum oscillations of magnetization (dHvA). According to the Onsager relation $F=(\phi{_0}/2\pi^2)=(\hbar/2\pi e)A$, the frequency F is proportional to the extreme cross section of the FS normal to magnetic field. Thus, dHvA oscillations is an effective method to study the electronic structures of topological materials, which can reveal the detailed information of the FS. Figure 2(a) presents the isothermal magnetization of a typical single crystal with B//[001] configuration, which exhibits evident dHvA oscillations. The oscillatory components of magnetization are obtained after subtracting a smooth background and displayed in Fig. 2(b). Twelve fundamental frequencies are extrcted from the FFT analysis. The low frequency about 2.1 T comes from the data processing and is not the intrinsic dHvA oscillations signal. The oscillatory components versus 1/B of dHvA oscillations can be described by the LK formula\cite{shoenberg2009magnetic}:

\begin{equation}\label{equ1}
\Delta M \propto -B^{1/2}\frac{\lambda T}{sinh(\lambda T)}e^{-\lambda T_D}sin[2\pi(\frac{F}{B}-\frac{1}{2}+\beta+\delta)]
\end{equation}
where $\lambda=(2\pi^2 k{_B} m^*)/(\hbar eB)$. $T{_D}$ is the Dingle
temperature. For the 2D system, $\delta=0$ and for the 3D system, $\delta=\pm1/8$.
$\beta=\phi_B/2\pi$ and $\phi_B$ is the Berry phase. The inset of Fig. 2(c) shows the normalized temperature-dependent FFT amplitudes
and the fitting by the thermal factor $R_T=(\lambda T)/sinh(\lambda T)$ in LK formula. The effective masses are estimated to be $m^*_{F_2}=0.05m_e$, $m^*_{F_3}=0.12m_e$, $m^*_{F_4}=0.10m_e$, $m^*_{F_5}=0.11m_e$, $m^*_{F_7}=0.08m_e$ and $m^*_{F_{11}}=0.15m_e$.

Figure 3(a) exhibits the calculated band structures of PtGa without SOC. There is spin-1 excitation at $\Gamma$ and double Weyl fermion at R in the first Brillouin zone with the Chern number $\pm$2, respectively. There is a ball-like electronic FS at $\Gamma$ with one FS's extreme cross section cut at k$_z$=0. The number of FS's extreme cross section at R cutting at k$_z$=$\pi$ is also one (Fig.3(b)-Fig.3(d)). When the SOC is included, the band is split which leads the  spin-1 excitation and double Weyl fermion evolve into spin-3/2 RSW fermion and TR doubling of the spin-1 excitation with the Chern number $\pm$4, respectively as shown in Fig. 4(a). Figure 4(b) displays the calculated FSs of PtGa with SOC in the first Brillouin zone as shown in inset of Fig. 4(e). The ball-like electronic FS at $\Gamma$ evolve into two concentric ball-like FSs with two FS's extreme cross section cutting at k$_z$=0. The number of FS's extreme cross section at R cut at k$_z$=$\pi$ is also two. Figure 4(c) exhibits FSs' cross section cutting at k$_z$=0. There are eight fundamental extreme cross sections. S$_5$, S$_6$, S$_7$ and S$_8$ at M point show a quarter of the FSs' extreme cross sections. Fig. 4(d) exhibits FSs' cross sections cutting at k$_z$=$\pi$. There are four fundamental extreme cross sections. S$_9$, S$_{10}$ show half of the FSs' cross section at M point and S$_{11}$, S$_{12}$ show  a quarter of the FSs' extreme cross sections at R point. The calculated frequencies originating from S$_1$ and S$_2$ are 13.0 T and 61.7 T corresponding to the observed F$_1$ (11.2 T) and F$_3$ (66.2 T), respectively.  The calculated  frequencies originating from 4S$_{11}$ and 4S$_{12}$ are 1699.0 T and 1987.6 T corresponding to the observed F$_{11}$ (1723.9 T) and F$_{12}$ (1937.3 T), respectively. As we can see in Fig. 4(c), the gaps between S$_5$ and S$_6$ along M-X, the S$_7$ and S$_8$ along M-X and M-$\Gamma$ are very small. The magnetic breakdown easily happens between S$_5$ and S$_6$, S$_7$ and S$_8$. The observed  frequency F$_9$ (333.7 T) may originates from the difference between 4S$_8$ and 4S$_7$ (346.8 T). The matching between the FSs' extreme cross sections and the observed frequencies is shown in Fig. 4(e). The originations of the rest observed frequencies may from the FSs' extreme cross sections cutting between k$_z$=0 and k$_z$=$\pi$ and/or the complicated magnetic breakdown.

\section{Summary}

In conclusion, we have successfully synthesized the single crystals of PtGa by Bi-flux method. It shows a metallic behavior at zero field. The large unsaturated MR reaching ~230\% at 1.8 K and 14 T with B//[111] has been observed, which decreases gradually with $\theta$ increasing from $\theta=0^o$ to $\theta=90^o$ and tends to saturate at $\theta=90^o$ under high field. The dHvA oscillations have been observed, from which the light cyclotron effective masses are extracted from the fitting of the thermal damping term by the LK formula, indicating the possible existence of massless Weyl Fermions. The dHvA oscillations also reveal that there are two concentric ball-like electronic FSs with two FSs' extreme cross sections at $\Gamma$ and two concentric circle-like FSs' extreme cross sections at R indicating the possible existence of spin-3/2 RSW fermion and TR doubling of spin-1 excitation which is in agreement with our first-principles calculations including SOC. ARPES is required for further study of its band structures and Fermi arcs in the surface states.

\emph{Note added.} When the paper is being finalized, we notice one related work, in which similar magneto-transport measurements are reported with ARPES confirming the topological characteristics in PtGa crystals grown with a different method\cite{yao2020observation}.

\section{Acknowledgments}

This work is supported by the Ministry of Science and Technology of China (2019YFA0308602, 2018YFA0305700 and 2016YFA0300600), the National Natural Science Foundation of China (No.11874422, No.11574391), the Fundamental Research Funds for the Central Universities, and the Research Funds of Renmin University of China (No.19XNLG18, No.18XNLG14). H.W. also acknowledges support from the Chinese Academy of Sciences under grant number XDB28000000, the Science Challenge Project (No.  TZ2016004), the K. C. Wong Education Foundation (GJTD-2018-01), Beijing Municipal Science \& Technology Commission (Z181100004218001) and Beijing Natural Science Foundation (Z180008).
\bibliography{Bibtex}
\end{document}